# Aging effects of dodecagonal quasicrystal formed in Mn-Cr-Ni-Si alloys

**Kotoba Toyonaga and Tsutomu Ishimasa**

Graduate School of Engineering, Hokkaido University, Kita-ku, 060-8628 Sapporo, Japan

E-mail: ishimasa@eng.hokudai.ac.jp

**Abstract** The formation of Mn-Cr-Ni-Si dodecagonal quasicrystal has been studied at 600 and 700 °C. The growth process of the quasicrystal from as-cast β-Mn phase was explained by Johnson-Mehl-Avrami equation. In order to evaluate structural quality of the quasicrystal, distortion in the diffraction pattern observed along the 12-fold axis was analyzed. This analysis indicated that the aging at 600 °C for 100 h is the best condition to synthesize the quasicrystal.

## 1. Introduction

A dodecagonal quasicrystal is a three-dimensional structure that is characterized by a unique 12-fold symmetry axis in the reciprocal space. Recently a new dodecagonal quasicrystal has been found in Mn-Cr-Ni-Si system, of which five-dimensional space group is $P12_6/mmc$ [1]. High resolution electron microscopy has revealed that the projection of the quasicrystal along the 12-fold axis is regarded as nonperiodic arrangement of a square and an equilateral triangle with the common edge length $a = 4.560$ Å [2]. Furthermore, geometrical analysis in the phason space has revealed the presence of an acceptance region that indicates the quasiperiodic order in a limited region within approximately 130 Å in diameter. This quasicrystal is precipitated from a β-Mn phase by heat treatment at 700 °C [1]. The purpose of the present study is to clarify aging effects on the formation of the quasicrystal.

## 2. Experimental procedures

At the composition $Mn_{72.0}Cr_{5.5}Ni_{5.0}Si_{17.5}$, alloy specimens were prepared using an arc furnace in an Ar atmosphere. The as-cast alloy just after melting included the cubic β-Mn type phase exclusively. They were further aged in an electric furnace at 600 or 700 °C for 57 - 611 h. The specimen enclosed in a silica ampoule was put into the furnace already set at each aging temperature beforehand. This procedure enables us to study the effect of the aging. Structural characterization was carried out using both a transmission electron microscope JEM200CS operated at 200 kV and a powder X-ray diffractometer RINT2200 using Cu Kα-radiation. In order to quantify both integral intensity and width of reflections, profile decomposition was carried out assuming pseudo-Voigt function for each reflection, where the contributions of $K\alpha_1$ and $K\alpha_2$ components were considered. Selected-area diffraction patterns were observed along the 12-fold axis using an aperture corresponding to approximately 2500 Å in diameter. The indexing scheme of a dodecagonal quasicrystal proposed in [3] is used throughout this report.

## 3. Results and Discussions

Figure 1 presents a powder X-ray diffraction pattern of the $Mn_{72.0}Cr_{5.5}Ni_{5.0}Si_{17.5}$ alloy that was aged at 600 °C for 611 h. The dodecagonal quasicrystal coexists with the dominant cubic β-Mn type phase with the lattice parameter 6.270 (1) Å. The dodecagonal quasicrystal exhibits broad peaks as indicated bars at the lower part in Figure 1a. The lattice parameters were estimated by modified Cohen's method to be $a$ = 4.558 (3) Å and $c$ = 4.620 (4) Å. Here the parameters $a$ and $c$ correspond to the common edge length of the tiles and the period along the 12-fold axis, respectively. In Figure 1b, result of peak decomposition is presented, in which there are four reflections of the dodecagonal quasicrystal.

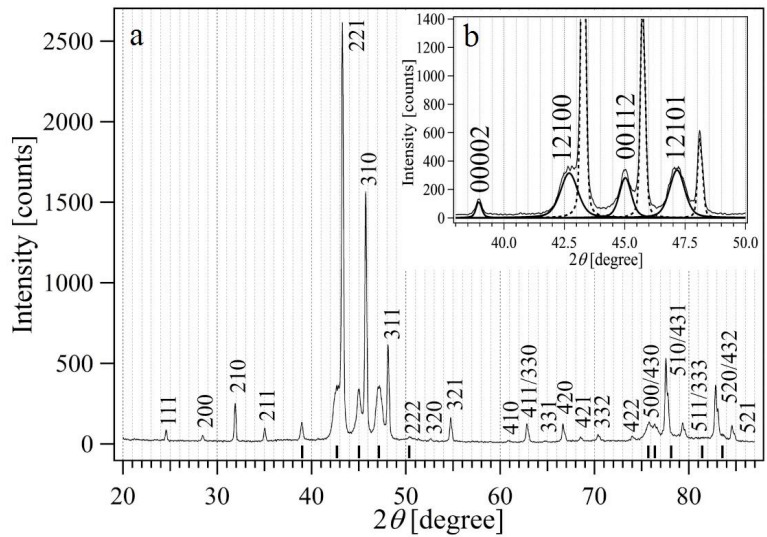
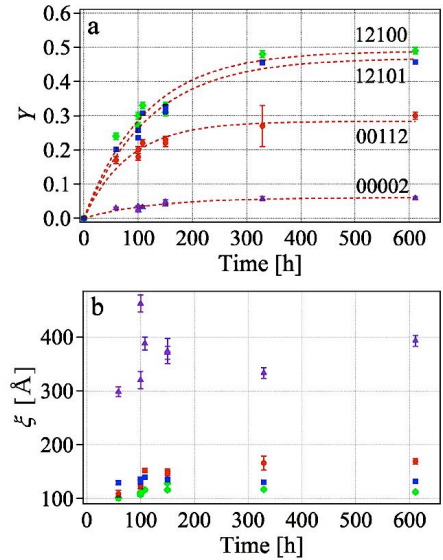

**Figure 1.** (a) Powder X-ray diffraction pattern of $Mn_{72.0}Cr_{5.5}Ni_{5.0}Si_{17.5}$. Indices of the β-Mn type phase are inserted. (b) Magnified part of the X-ray diffraction pattern. Four reflections can be detected: 00002, 12100, 00112, and 12101.

**Figure 2.** Effect of aging at 600 °C. (a) Normalized intensity, (b) Correlation length determined by the four reflections.

Figure 2 shows the growth of the quasicrystal during the aging at 600 °C. In Figure 2a, normalized intensities, $Y = I_g/I_{221}$, of the following four reflections are presented: $g$ = 00002, 12100, 00112, 12101. For the normalization, the intensity of the 221 reflection of the β-Mn type phase was used as a standard. Strictly speaking, the amount of the β-Mn type phase was decreased during the aging, but the decrease was small enough for this analysis. The growth process is well explained by Johnson-Mehl-Avrami Equation: $Y = Y_0(1-\exp(-Kt^n))$. Here $t$ [h] is aging time, and $Y_0$ is constant. $K$ is the rate parameter. When the exponent $n$ = 1 was used, the excellent agreement shown in Figure 2a was obtained. The parameter $K$ is approximately 0.010 $h^{-1}$ for all the reflections. Similar growth was observed also for aging temperature 700 °C with larger $K$, approximately 0.020 $h^{-1}$. These results indicate gradual growth of the quasicrystal during the aging, and suggest stability or metastability of the quasicrystal at these temperatures. Figure 2b presents evolution of the correlation length $\xi$ [Å] determined from the peak width $\Delta\theta$, full width at half maxima, using the following equation: $\xi = \lambda/2\Delta\theta\cos\theta$. Here $\lambda$ and $\theta$ denote the wavelength of the X-ray and the Bragg angle, respectively.

Correlation lengths were slightly increased during the period between 50 and 150 h, and reached to each final value. The correlation length along the quasiperiodic direction is approximately 100 Å, but that along the periodic direction is much larger. (Notice the difference between 12100 and 00002 reflections.) Similar behavior was also observed at 700 °C. The correlation length along the quasiperiodic direction agrees with the domain size observed in the electron micrograph [2].

Examples of electron diffraction patterns are presented in Figure 3. The 12-fold symmetric pattern in Figure 3a was observed in the specimen A aged at 600 °C for 100 h, where the suffix "A" denotes one of two specimens synthesized under the same aging condition. All the reflections are indexed by the indexing scheme with $h_5 = 0$ [3], where the lattice parameters are $a = 4.558$ Å and $c = 4.620$ Å. Both spot-like shapes of reflections and weak diffuse scattering indicate high structural quality of this quasicrystal. However, deviations from the ideal pattern were frequently observed. In Figure 3b, there is diffuse scattering. The pattern in Figure 3c shows 6-fold like symmetry rather than 12-fold.

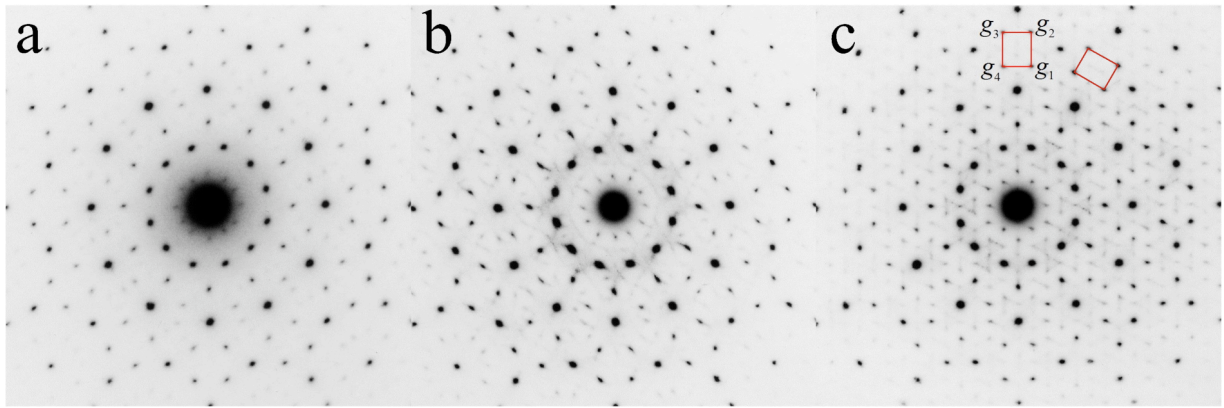

**Figure 3.** Electron diffraction patterns. (a) Nearly perfect pattern with small standard deviations, $\sigma(\beta) = 0.021$ and $\sigma(\mu) = 0.039$. (b) Pattern with diffuse scattering observed in the specimen aged at 600 °C for 611 h. $\sigma(\beta) = 0.024$ and $\sigma(\mu) = 0.030$. (c) Pattern showing 6-fold like symmetry observed in the specimen aged at 700 °C for 130 h. $\sigma(\beta) = 0.19$ and $\sigma(\mu) = 0.094$.

In order to evaluate the quality of quasicrystals, two-dimensional positional data of reflections were measured in each electron diffraction pattern. We are interested in the shape of the quadrangle formed by the four reflections, $g_1 = \bar{1}0320$, $g_2 = \bar{1}0330$, $g_3 = \bar{2}0330$ and $g_4 = \bar{2}0320$. They form a square in the ideal case with the edge length $a^* = 0.1267$ Å$^{-1}$. Actually the vertical angles of the quadrangles were measured to be 90.0° with the standard error typically 0.2°. Then they are regarded as a rectangle in the diffraction patterns observed in this study. The rectangle has two edges, $a_x^* = |g_1 - g_4|$ and $a_y^* = |g_3 - g_4|$. Two parameters are introduced here: $\beta = a_y^*/a_x^*$ and $\mu = a_x^*/\overline{a^*}$ (or $\mu = a_y^*/\overline{a^*}$). Here $\overline{a^*}$ denotes the average of 48 edge lengths of 12 rectangles experimentally measured. The parameters $\beta$ and $\mu$ express rectangularity and normalized edge length, respectively. The standard deviations of these parameters, $\sigma(\beta)$ and $\sigma(\mu)$, were calculated for 12 values of $\beta$ and 48 values of $\mu$ experimentally measured, respectively. Both standard deviations are zero in the ideal case, if there is no experimental error. With respect to these standard deviations, it may be useful to describe two typical cases. One is the hexagonal approximant $Mn_{70}Cr_{10}Si_{20}$ with the lattice parameters $a = 16.985$ Å and $c = 4.625$ Å [4]. In this case, $\beta = 2/\sqrt{3}$ or $\sqrt{3}/2$, and the standard deviations are $\sigma(\beta) = 1/4\sqrt{3} \approx 0.144$, $\sigma(\mu) = 1 - \sqrt{3}/2 \approx 0.134$. The other is the tetragonal approximant, σ-phase, with the lattice parameters $a = 8.806$ Å and $c = 4.624$ Å. In this case, there are two kinds of squares with different edge lengths. The standard deviations are σ(β) = 0 and σ(μ) = (√2-1)² ≈ 0.172. In Figure 4, the two kinds of standard deviations are presented for the four specimens: A and B aged at 600 °C for 100 h and that aged at 600 °C for 611 h, and at 700 °C for 130 h. For this analysis, seven, five, nine, and twelve data sets were used respectively for each specimen. The specimen A aged at 600 °C for 100 h exhibits remarkably small deviations with respect to both β and μ. However, other specimens show larger deviations and wider distributions in σ(β) and σ(μ). These results indicate tendency of structural fluctuation caused by longer aging time or higher aging temperature. Then the dodecagonal quasicrystal formed in this Mn-based alloy is possibly a metastable phase, while the growth behavior is well explained by Johnson-Mehl-Avrami Equation. The specimen B at 600 °C for 100 h shows larger standard deviations than the specimen A, and this difference clearly indicates the limitation of the present experiment.

The shift of reflections from the exact symmetric positions can be related to the tile rearrangement in the direct space that is called phason disorder [5]. The relationship between the standard deviations described here and the phason tensor is the subject to be studied.

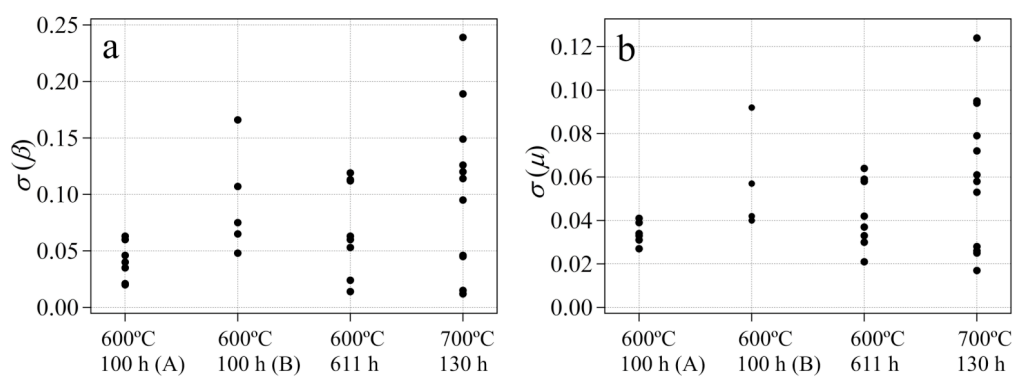

**Figure 4.** Deviations from the exact 12-fold symmetry showing high-quality of the specimen A aged at 600 °C for 100 h. (a) Standard deviation σ(β). (b) Standard deviation σ(μ).

## 4. Conclusion

In the $Mn_{72.0}Cr_{5.5}Ni_{5.0}Si_{17.5}$ alloy, the dodecagonal quasicrystal is slowly grown from the matrix of β-Mn type phase by aging at 600 or 700 °C. This growth process is explained by Johnson-Mehl-Avrami Equation. However, the detailed analysis of electron diffraction patterns suggested the increase of the deviations from the ideal dodecagonal quasicrystal during the aging. In the limit of this experiment, the best condition to synthesize the dodecagonal quasicrystal is aging at 600 °C for 100 h. Even under this condition, the correlation length in the quasiperiodic direction is limited within 100 Å. Synthesis of the dodecagonal quasicrystal with high quality and uniformity is still beyond our control.

**Acknowledgement**
The authors thank Shuhei Iwami for his support in the specimen preparation.